# Persistence of Induced Wormholes in Brane Worlds

Enrico Rodrigo
*Department of Physics and Astronomy, University of California at Irvine,
Irvine, CA 92697-4575, USA*
and
*Department of General Studies, Charles Drew University
Los Angeles, CA 90059, USA*

Higher-dimensional black holes have long been considered within the context of brane worlds. Recently, it was shown that the brane-world ethos also permits the consideration of higher-dimensional wormholes. When such a wormhole, preexisting in the bulk, impinges upon our universe, taken to be a positive-tension 3-brane, it induces the creation in our universe of a wormhole of ordinary dimensionality. The throat of this wormhole might fully constrict, pinch off, and thus birth a baby universe. Alternatively, the induced wormhole might persist. I show that persistence is more likely and note that the persistent wormhole manifests as a particle-like object whose interaction with cosmic matter is purely gravitational. As such it may be considered a viable candidate for the sort of weakly interacting massive particle long believed to be the prime constituent of cold dark matter.

## 1. Introduction

Wormholes have been discussed in the context of brane worlds ever since it was noticed [1] that the original Randall-Sundrum two-brane construction [2] meets the formal definition of a wormhole. In this construction the branes served as boundaries of a higher-dimensional spacetime – the bulk. This idea of branes as boundaries can be extended to bulk spacetimes that are not simply connected. In this case the bulk would contain holes that are bounded by closed branes. These holes would be regions of literal nothingness or void. The closed branes serving as boundaries between the bulk and this void may be modeled as the throats of thin-shell semi-wormholes [3].

A bulk containing such wormholes in addition to the brane defining our universe is not dissimilar to the oft-considered brane worlds featuring a bulk inhabited by one or more black holes. Unlike the latter case, which implies the existence of singularities in the bulk, a brane world complimented by bulk-dwelling wormholes faithfully adheres to the canonical proscription of off-brane matter. For this reason the interaction of wormholes in the bulk with the brane defining our universe is at least as interesting a priori as the analogous interaction between our universe and bulk-dwelling black holes. The latter, which has been the subject of recent investigations [4, 5], is complicated by the existence of the event horizon of the black hole. Frolov's recent model of this interaction shows that the geometry of the brane is increasingly distorted by the approach of a higher-dimensional black hole, until it induces within the brane the formation of new black hole, whose dimensionality is lower – matching that of the brane. This induced black hole forms, when the brane enters the horizon of the original bulk black hole.

Recently, it was suggested [6] that the wormhole-brane interaction is analogous to that between a brane and a black hole, with the role of the black hole's horizon being played by the wormhole's throat. The existence of a throat would seem to permit the envelopment of the incident wormhole by the brane on which it has impinged. The result of this envelopment, analogous to the aforementioned denouement of certain brane-black-hole encounters, would be the induced formation within the brane of a new wormhole. The purpose of this note is to consider the result of an encounter between a brane and a wormhole. I shall in particular address the question of whether *partial* envelopment of the wormhole by the brane – the condition corresponding to a persistent induced wormhole – necessary proceeds to *total* envelopment. The latter is tantamount to the birth of a baby universe that occurs, when the throat of the induced wormhole becomes arbitrarily small.

## 2. Brane Description

In order to determine whether envelopment of a wormhole by a brane is in fact possible, we will consider the action of the static brane interacting gravitationally with a bulk wormhole. From the point of view of the bulk, the wormhole-brane system is not spherically symmetrical. Gravitational waves will therefore be emitted, as the wormhole impinges on the brane. This dissipative effect suggests that the configurations of the brane that minimize its static action are possible end states of a dynamic encounter. Such an encounter would in general require numerous bounces of the wormhole against the brane, before the final configuration predicted by the static action is reached. I shall make no attempt to model the dynamics of the encounter or to estimate the rate at which the local energy density is dissipated by gravitational waves. Nor shall I consider explosive or otherwise dissipative effects of brane-brane interaction through the emission within the macro brane (identified with our universe) of outbound fluxes of standard-model fields. Rather, I will focus on whether a conservatively defined static action permits total envelopment of the bulk wormhole by the brane. If it does not, we may then conclude that an induced wormhole persists -- that partial envelopment is not necessarily an intermediate step toward total envelopment and the formation of a baby universe.

We begin our detailed description of the static result of a wormhole-brane encounter by specifying the bulk wormhole. Let it be an *N*-dimensional, asymptotically anti-de Sitter Reissner-Nordstrøm black hole spacetime from which a hyper-cylindrical region, centered on *r=0* and enclosed by "surface" with topology $S^{N-2} \times R^1$, has been excised. That is to say, each spacelike slice (obtained by holding the time coordinate constant) will be missing a central region enclosed by an (*N-2*)-sphere. By choosing the radius of this sphere to exceed that of the event horizon of the corresponding (nonrotating) black hole, we ensure that the resulting geometry describes a (horizon-free) semi-wormhole. At the boundary (with topology $S^{N-2} \times R^1$) between this spacetime and the void created by the aforementioned excision – this boundary being an (*N-2*)-sphere in each *t=constant* spacelike slice -- we place the world tube of a closed negative-tension brane. This is in effect a spherically symmetrical thin-shell semi-wormhole, whose metric has the form

$$ds^2 = g_{\mu\nu}dX^\mu dX^\nu = -F(r)dt^2 + \frac{dr^2}{F(r)} + r^2 d\Omega_{N-2} \qquad (1)$$

with the restrictions $r \geq r_T$ and $r_T > r_H$, where $r_T$ is the radius of the semi-wormhole's throat and $r_H$ is the root of $F(r) = 0$ that corresponds to the external event horizon of the relevant black hole, and $d\Omega_{N-2}$ is the usual "surface area" element of an $(N-2)$-dimensional unit sphere. For an $F$ corresponding to an asymptotically anti-de Sitter Reissner-Nordstrøm metric, a Wheeler-DeWitt-style treatment of the corresponding thin-shell semi-wormholes suggests that discrete radii exist at which these wormholes are quantum mechanically stable [3, 6, 11]. Envelopment by the brane constituting our universe of these stable, bulk-inhabiting, micro semi-wormholes – also known as "void bubbles" – would induce in our universe the formation of micro wormholes that would be all but indistinguishable from weakly interacting massive particles.

Let the dimensionality of the brane be $D$, where $D < N$. The coordinates $X^\mu$ of the $N$-dimensional bulk are given by $(X^\mu)=(t, r, \theta_1, \ldots \theta_{N-2})$ and those $\zeta^a$ of the $D$-dimensional brane by $(\zeta^a)=(t, s, \theta_1, \ldots \theta_{D-2})$. The embedding of the spherically symmetric brane within the bulk can be given in terms of the bulk coordinates parametrized by the radial brane coordinate $s$,

$$\begin{aligned} r &= r(s) \\ \theta_{D-1} &= \theta(s) \\ \theta_D &= \ldots = \theta_{N-2} = \pi/2 \end{aligned} \qquad (2)$$

Here we deviate slightly from Frolov's treatment [4] (whose notation I have adopted) in order to permit arbitrary spherically symmetric brane configurations, a large class of which his chosen parametrization cannot describe. The metric $\gamma_{ab}$ induced on the brane by the bulk's geometry,

$$\gamma_{ab} = g_{\mu\nu}\frac{\partial X^\mu}{\partial \zeta^a}\frac{\partial X^\nu}{\partial \zeta^b}, \qquad (3)$$

has by the definitions of the coordinate systems the line element

$$ds^2 = \gamma_{ab}d\zeta^a d\zeta^b$$
$$= -Fdt^2 + \left(F^{-1}r'^2 + r^2\theta'^2\right)ds^2 + r^2\sin^2\theta d\Omega_{D-2} \qquad (4)$$

where $r' \equiv dr/ds$ and $\theta' \equiv d\theta/ds$. Assuming the motion of the brane to be determined by the Dirac-Nambu-Goto action [7,8,9]

$$S = \int d^D\zeta \sqrt{-\det(\gamma_{ab})}, \qquad (5)$$

we see that in the static case the action becomes

$$S = \Delta t \Omega_{D-2} \int r^{D-2} \sin^{D-2}\theta \sqrt{r'^2 + Fr^2\theta'^2}\, ds \tag{6}$$

where $\Omega_{D-2}$ is the "surface area" of a (*D-2*)-dimensional unit sphere and $\Delta t$ is an arbitrary time interval. This yields the Euler-Lagrange equations

$$Jr'r^{D-2}\sin^{D-3}\theta = 0 \tag{7}$$

$$J\theta'r^{D-2}\sin^{D-3}\theta = 0 \tag{8}$$

where

$$\begin{aligned}J \equiv &-(D-2)r'^3\cos\theta - (D-2)r'Fr^2\theta'^2\cos\theta + (D-2)r'^2 rF\theta'\sin\theta \\ &+ (D-2)r^3 F^2\theta'^3\sin\theta - Fr^2\theta'r''\sin\theta + \tfrac{1}{2}Fr^4\theta'^3\frac{dF}{dr}\sin\theta \\ &+ F^2 r^3\theta'^3\sin\theta + \frac{dF}{dr}r'^2 r^2\theta'\sin\theta + 2Fr\theta'r'^2\sin\theta + Fr^2\theta''r'\sin\theta\end{aligned} \tag{9}$$

### 3. Wormhole Envelopment

The question of whether total envelopment of the bulk wormhole by the brane is a possible end state of a wormhole-brane encounter is now readily answered. Defining total envelopment by

$$r = a \tag{10}$$

$$\theta = \pi - \frac{s}{a} \quad , \tag{11}$$

where *a* is the throat radius of the incident bulk wormhole, we see that eq. (7) is immediately satisfied and that eq. (8) requires that $J = 0$. This condition becomes, after inserting eqs. (10) and (11) into (9),

$$\left.\frac{dF}{dr}\right|_{r=a} = -\frac{2(D-1)}{a}F(a). \tag{12}$$

Total envelopment also assumes that the throat of the wormhole induced in the brane has fully constricted to a filament connecting the brane to a spherical pocket universe that surrounds the throat of the bulk wormhole. This filamentary throat may be described by the equation $\theta = 0$ for $r > a$, which clearly satisfies the Euler-Lagrange equations (8) and (9). Because topology change is forbidden within general relativity, I am assuming that the existence of a filamentary throat (i.e. one whose radius is arbitrarily small) signals the presence of a topology change in whichever more permissive and presumably truer theory general relativity serves as a low-energy limit.

We require the bulk to be free of matter except at its brane boundaries – the throats of any higher-dimensional bulk semi-wormholes. Accordingly, we form these semi-wormholes by terminating higher-dimensional black hole solutions at a radius $a$ outside of its event horizon. Choosing a spherically symmetric black hole solution consistent within an anti-de Sitter bulk, we have

$$F = 1 - \frac{M_N}{r^{N-3}} + \frac{Q_N^2}{r^{2N-6}} - \Lambda_N r^2 \qquad (13)$$

where

$$M_N \equiv \frac{16\pi G_N M}{(N-2)c^2 \Omega_{N-2}} \qquad (14)$$

$$Q_N^2 \equiv \frac{k_N Q^2 G_N}{2(N-2)(N-3)} \left(\frac{8\pi}{\Omega_{N-2} c^2}\right)^2 \qquad (15)$$

$$\Lambda_N \equiv \frac{2\Lambda}{(N-1)(N-2)} \qquad . \qquad (16)$$

$M$, $Q$, $\Lambda$ the asymptotically observed mass, charge, and cosmological constant. $G_N$ and $k_N$ are respectively the $N$-dimensional gravitational and electrostatic constants, and $\Omega_{N-2}$ is the "surface area" of an ($N$-2)-dimensional unit sphere. Inserting eq. (13) into (12) and specializing to the case of a 3-brane universe ($D$ = 4), we have

$$-\Lambda_N a^{2N-4} + \tfrac{3}{4} a^{2N-6} + \frac{N-9}{8} M_N a^{N-3} - \frac{N-6}{4} Q_N^2 = 0 \qquad (17)$$

which becomes in the case of an uncharged bulk wormhole,

$$-\Lambda_N a^{N-1} + \tfrac{3}{4} a^{N-3} + \frac{N-9}{8} M_N = 0 \qquad (18).$$

Comparing eq. (18) with the equation for the horizons of the bulk wormhole – namely, $F = 0$, i.e.

$$-\Lambda_N r^{N-1} + r^{N-3} - M_N = 0 \quad , \qquad (19)$$

we find that the positive real roots of (18) (for the values of $N$ at which they exist) are necessarily smaller than those of (19), the latter corresponding to black hole horizons. In other words, total envelopment is only possible if the throat radius $a$ of the bulk wormhole is smaller than the event horizon of the corresponding black hole. Total envelopment by the brane of an incident bulk wormhole requires the wormhole to be a black hole.

If a bulk wormhole is not a black hole, its envelopment, then, must only be partial. The encounter of the brane with a bulk wormhole must therefore induce the formation within the brane of the structure corresponding to partial envelopment -- a wormhole whose dimensionality matches that of the brane. This wormhole persists in the sense that it presumably remains in the state of partial envelopment and does not, for the reason adduced, become fully enveloped and pinch off. Were it macroscopic, denizens of the brane would recognize it as a wormhole to a pocket universe. It is microscopic, however, with radius perhaps on the order of $10^{-22}$ cm and a mass perhaps on the order of $10^4$ TeV [6]. Hence, such an induced wormhole would be perceived instead as an ultra-massive particle, whose interactions are purely gravitational. This sort of weakly interacting massive particle-like object could serve as a constituent of dark matter (see [10] for a recent review of other dark matter candidates). To ensure a sufficient quantity of these WIMP-like objects, one might suppose the bulk to be awash in tiny semi-wormholes, each of whose throats is coincident with an (*N*-2)-spherical micro brane. Certain of these void bubbles would impinge upon the macro brane that constitutes our universe, become embedded there, and manifest as tiny wormholes that are perceived by us brane-dwellers as the aforementioned WIMPs.

## 4. Conclusion

To summarize, horizon-free bulk wormholes that encounter the brane presumed to constitute our universe cannot become totally enveloped by it. The encounter will not, therefore, result in the birth of a baby universe. Instead, envelopment by the brane will be partial and will thereby manifest as induced wormholes that persist. Because of their microscopic scale and purely gravitational interactions, these wormholes will be perceived as particles of dark matter.

## REFERENCES

[1] L. Anchordoqui and S. Perez Bergliaffa, *Phys. Rev.* **D62**, 067502 (2000)
[2] L. Randall and R. Sundrum, *Phys. Rev. Lett.* **83**, 3370 (1999)
[3] C. Barceló and M. Visser, *Nucl. Phys.* **B584**, 415 (2000)
[4] V. Frolov, *Phys. Rev. D* **74**, 044006 (2006)
[5] V. Frolov, M. Snajdr, D. Stojkovic, *Phys. Rev. D* **68**, 044002 (2003)
[6] E. Rodrigo, *Phys. Rev. D* **74**, 104025
[7] Y. Nambu, in *Lectures at the Copenhagen Symp. on Symmetries and Quark Models*, p.269, Gordon and Breach Book Co., New York (1970)
[8] P. Dirac, *Proc. Royal Soc. London* **A268**, 57 (1962)
[9] T. Goto, *Prog. Theor. Phys.* **46**, 1560 (1971)
[10] G. Bertone, D. Hooper, J. Silk, *Phys.Rept.* **405**, 279 (2005)
[11] M. Visser, *Phys. Rev. D* **43**, 402 (1991)